# Antibacterial melamine foams decorated with *in situ* synthesized silver nanoparticles

*Javier Pinto[a]\*, Davide Magrì[a], Paola Valentini[b], Francisco Palazon[c], José A. Heredia-Guerrero[a], Simone Lauciello[d], Suset Barroso-Solares[a], Luca Ceseracciu[e], Pier Paolo Pompa[b], Athanassia Athanassiou[a], and Despina Fragouli[a]\** a: Smart Materials, Istituto Italiano di Tecnologia, Via Morego 30, 16163 Genova, Italy. Email: jpintosanz@gmail.com (J. Pinto), despina.fragouli@iit.it (D. Fragouli)

b: Nanobiointeractions & Nanodiagnostics, Istituto Italiano di Tecnologia, Via Morego 30, 16163 Genova, Italy

c: Nanochemistry Department, Istituto Italiano di Tecnologia, Via Morego 30, 16163 Genova, Italy

d: Electron Microscopy Facility, Istituto Italiano di Tecnologia, Via Morego 30, 16163 Genova, Italy

e: Materials Characterization Facility, Istituto Italiano di Tecnologia, Via Morego 30, 16163 Genova, Italy

**Abstract**

A new and straightforward single-step route to decorate melamine foams with silver nanoparticles (ME/Ag) is proposed. Uniform coatings of silver nanoparticles with diameters less than 10 nm are formed *in-situ* directly on the struts surface of the foams, after their dipping in an $AgNO_3$ solution. We prove that the nanoparticles are stably adhered on the foams, and that their amount can be directly controlled by the concentration of the $AgNO_3$ solution and the dipping time. Following this production route, ME/Ag foams can be obtained with silver content ranging between 0.2 to 18.6 wt.% and excellent antibacterial performance, making them appropriate for various applications. Herein we explore the possibility to use them as antibacterial filters for water treatment, proving that they are able to remove completely *Escherichia coli* bacteria from water when filtered at flow rates up to 100 mL/h·cm$^2$ due to the release of less than 1 ppm of $Ag^+$ ions by the foams. No bacterial regrowth was observed after further dilution of the treated water, to arrive below the safety threshold of $Ag^+$ for drinking water (0.1 ppm), demonstrating the excellent bactericide performance of the ME/Ag filters.

# 1. Introduction

According to the World Health Organization (WHO), contaminated drinking water is estimated to cause half a million of deaths each year, whereas *ca.* 1.8 billion people consumed drinking water sources contaminated with feces in 2016.[1] Sanitation of water destined for human use is, therefore, an issue of utmost importance, requiring the development of new suitable and efficient approaches able to remove and inactivate pathogenic microorganisms. For such applications, in the last years, the use of silver nanoparticles (Ag NPs) has arisen as an effective dinsifectant agent.[2–7] However, the direct use of Ag NPs in the drinking water presents some practical drawbacks, such as their possible aggregation, which could decrease their antibacterial efficiency, and their potential toxicity for living organisms.[8,9] With the aim to overcome these problematics the Ag NPs have been immobilized on different porous substrates such as paper,[10] natural fibers,[11] ceramic filters,[12,13] and polymer foams,[8,14–17] making thus possible their use in water treatment applications. Among them, polymer foams offer several advantages, presenting high specific surface area permitting to the immobilized Ag NPs to directly interact with the water,[8,14–17] while their porous structure enables their use in filtration setups.[15,16]

Latex, polyurethane, and melamine (ME) foams have been proposed so far for the development of Ag loaded antibacterial foams for water treatment,[8,14–17] formed by the direct transfer of previously synthesized Ag NPs on the foams' surface by dip-coating,[14–16] or by mixing of the NPs with the polymer matrix before the formation of the foams.[17] To achieve a satisfactory antibacterial performance, the dip coating processes are better choice than the mixing, since they allow the NPs to be exposed onto the foams' surface whereas after mixing the NPs are covered by polymer layers. The data obtained by the filtration processes proposed in some works[15,16] lack information on the possibility to utilize these materials for the sanitation of drinking water, as no antibacterial results were presented for concentrations of $Ag^+$ ions below the safety threshold established by the WHO for drinking water (0.1 ppm).[4] Moreover, the dip-coating procedures employed do not always ensure the NPs' stable anchoring on the foams, and thus, do not exclude the possibility of their release in the treated water. Finally, the efficiency of the methods to transfer silver to the foams was either not studied,[16] or was found to be very low (~ 0.3 wt.%).[15]

The best route to stably link sufficiently high amounts of Ag NPs onto the porous substrates is to form the NPs directly onto the solid, following an *in situ* synthesis approach.[8,10,11,18–22] This procedure requires the reduction of a silver precursor (usually $AgNO_3$) using chemical or physical processes in order to form Ag NPs. Specifically, it has been demonstrated that the utilization of $NaBH_4$,[8,10,23,24] glucose,[11] polydopamine,[18] ethylene glycol,[19] or the UV-induced photo-reduction[20–22] of the silver precursor results in the formation of well-defined Ag NPs on solid surfaces.

So far, for water treatment applications using polymer foams, this process was followed utilizing ME foams as substrates where the Ag NPs have been formed *in situ* using a multiple-step procedure.[8] In particular, ME foams are decorated with Ag NPs and graphene oxide (GO) by the sequential treatment with acetone, dopamine hydrochloride solution, GO/$Ag^+$ solution, and finally with a $NaBH_4$ solution. The maximum amount of silver transferred on the foams was 9.8 wt.%, while their antibacterial performance was demonstrated exclusively in the liquid absorbed by the foam after being placed in a bacterial suspension.[8] The effect of the treated foams on the remaining bacterial suspension was not tested, while no information was presented on their potential use in filtration setups, necessary for water treatment applications.

Herein, we propose an alternative route for the fabrication of antibacterial ME foams, employing a single-step *in situ* synthesis of Ag NPs into the surface of the struts of ME foams. The proposed procedure takes advantage of the R–$NH_2$ groups present in the ME foams, which are known to induce the chemical reduction of the $AgNO_3$ metal precursor, resulting in the formation of Ag NPs in other systems.[25,26] The proposed procedure allows a fine control of the silver amount transferred to the ME foams, reaching a higher Ag content compared to the so far reported Ag-foam composite systems. At the same time, the porous structure and mechanical properties of the ME foams are not affected by the functionalization procedure, which makes these foams suitable for water treatment. The obtained foams are able to release $Ag^+$ ions in the polluted water and provide efficient antibacterial performance even using a continuous-flow setup, allowing their use in filtration applications. Moreover, the type of antibacterial performance provided by these foams (bacteriostatic or bactericide) is investigated in detail, a relevant issue usually not addressed by previous works, as highlighted by the WHO.[4] We prove that the obtained foams act as bactericidal filters, avoiding any regrowth of the bacteria in the filtered media even when the $Ag^+$ concentrations are below the safety threshold for drinking water (0.1 ppm).

## 2. Experimental procedures

2.1 Materials

Melamine (ME) foams (commercial grade Basotec® G+, BASF) were kindly provided by LAPE HD Srl (Empoli, Italy). These foams present a very high porosity (0.994), open pores, and a broad pore size distribution with a mean pore size below 500 μm. Silver nitrate ($AgNO_3$) was purchased in the form of white powder by Sigma-Aldrich (product code: 209139). Tetrahydrofuran (THF) was purchased from Sigma-Aldrich and used as a solvent. Distilled water was employed as a co-solvent as well as for both the $Ag^+$ ions and the NPs release tests.

2.2 Treatment procedure of the ME foams

*In situ* synthesis of Ag NPs on the surface of the struts of ME foams was achieved as follows: Solutions with different $AgNO_3$ concentrations (0.05, 0.10, 0.85, and 1.70 mg/mL) were prepared in a $H_2O$:THF (1:19 v:v) solvent mixture. Then, ME foam samples with dimensions 2 x 1 x 1 $cm^3$ were immersed in 20 mL of the prepared solutions and kept under shaking for a certain time (ranging from 1 to 7 days). Once the desired reaction time was reached, the samples were extracted by the precursor solution, and were subsequently subjected to 5 washing cycles. In each washing cycle the samples were immersed in 20 mL of distilled water and kept under continuous shaking for 2 minutes. These steps were necessary in order to remove any remaining THF solvent, unreacted precursor and non-stably linked Ag NPs on the foams. As proved by the UV-vis spectroscopy study of the liquid of the fifth washing cycle, no THF, $AgNO_3$ or Ag NPs were released by the functionalized foam (see Supporting Information, Figure S1). After the rinsing, the samples were left to dry overnight prior to further characterization.

2. 3 Experimental techniques

UV-vis spectroscopy

UV-vis spectra of the $AgNO_3$ solutions in $H_2O$:THF (1:19, v:v) and of the rinsing residues were obtained with a Varian Cary 6000i UV-visible-NIR spectrophotometer in double beam configuration using $H_2O$:THF (1:19, v:v) mixture or distilled water, respectively, as reference samples. All spectra were recorded in the range from 250 to 800 nm. The presence of Ag NPs was determined by their characteristic plasmon peak at about 400-450 nm.[17,26]

X-ray photoelectron spectroscopy (XPS)

XPS was carried out with an Axis Ultra DLD spectrometer under $10^{-9}$ mbar pressure. Monochromatic Al Kα source with photon energy 1486.6 eV was used with an emission current of 20 mA and an operating voltage of 15 kV. High-resolution spectra were acquired with a step of 0.1 eV and an analyzer pass energy of 10 eV. Surface charging was neutralized with low-energy electrons (4 eV) and energy calibration was performed by setting the C-C / C-H component of the C 1s spectrum to a fixed binding energy value of 284.5 eV. Data analysis was performed with CasaXPS software.

Attenuated Total Reflection-Fourier Transform Infrared (ATR-FTIR) spectroscopy

Infrared spectra were obtained with an ATR accessory (MIRacle ATR, PIKE Technologies) coupled to an FTIR spectrometer (Equinox 70 FT-IR, Bruker). All spectra were recorded in the range from 4000 to 600 cm$^{-1}$ with a resolution of 4 cm$^{-1}$, accumulating 128 scans.

High-Resolution Scanning Electron Microscopy (HRSEM)

The porous structure of the ME foams and of the surface of their struts were characterized by HRSEM microscopy using a JEOL JSM-7500LA (JEOL, Tokyo, Japan) equipped with a cold field emission gun (FEG), operating at 10 kV acceleration voltage. The electrons used for imaging were backscattered electrons. Energy Dispersive Spectroscopy (EDS) was employed to identify the presence and distribution of silver on the surface of the struts. Although silver aggregates were identified by HRSEM in the surface of the struts it was not possible to identify individual Ag NPs by this technique.

Transmission Electron Microscopy (TEM)

The presence of Ag NPs on the struts surface of the ME foams was confirmed by TEM observations. Samples were firstly infiltrated in a low viscosity Spurr (SPI-Chem) epoxy resin. Once the resin has cured for 48h in oven at 65°C, 70 nm thick sections were cut with a Leica EMU C6 ultra-microtome. Transmission electron microscopy (TEM) images were collected with a Jeol JEM 1400 Plus (Jeol, Japan) electron microscope (Electron Microscopy Facility – Fondazione Istituto Italiano di Tecnologia), operating at an acceleration voltage of 120 kV, and recorded with a 11 Mp fiber optical charge-coupled device (CCD) camera (Gatan Orius 830). Size distribution of the Ag NPs was determined using Image J/FIJI.[27]

Inductive couple plasma-optical emission spectroscopy (ICP-OES)

The silver content in the foams, and the release of the Ag$^+$ ions when the foams are in contact with distilled water were evaluated using an ICP-OES spectrometer (iCAP 6300, Thermo).

In the first case, for the analysis of the silver content in foams, a microwave digestion system (MARS Xpress, CEM) was used to digest 5 mg of foams in 2.5 mL of nitric acid (70%, Sigma-Aldrich). The solid degradation reaction was performed at 180° C for 15 minutes. Before the analysis, the samples were diluted with milliQ water up to 25 mL and filtered through PTFE syringe filters (15 mm, pore size 0.45 µm, Sartorius).

In the second case, the $Ag^+$ ions release of the foams was determined by placing pristine and Ag treated ME (ME/Ag) foams in distilled water (2 mg of foam per mL of distilled water) under continuous shaking at 200 rpm for time intervals ranging between 0.5 and 48 h. After each time interval, the foams were extracted and the remaining liquid was analyzed. In particular, every solution sample collected (250 µL) was previously treated with 2.5 mL of nitric acid (70%, Sigma-Aldrich) overnight at room temperature. Then the digested samples were diluted up to 25 mL with milliQ water and, before the analysis, were filtered through polytetrafluoroethylene (PTFE) syringe filters (diameter 15 mm, pore size 0.45 µm, Sartorius).

The same method of analysis was followed also in the case of the evaluation of the $Ag^+$ ions released by the foam during the filtration experiments. In particular, aliquots of 250 µL from the water portions of 1 mL collected at random times during the filtration process, were collected and analyzed as previously described.

Antibacterial tests

The use of the ME/Ag foams for water treatment was first analyzed by tube tests in which 2 mg of the ME or ME/Ag foams were placed in 1 mL of *Escherichia coli* (*E. coli*) suspension (8,000,000 colony forming units (CFU)/mL *Luria Bertani* medium) and kept under shaking at 200 rpm for 4 hours. Bacterial growth after 30, 60, 90, 120, 150, 180, 210, and 240 minutes of interaction with the foams was determined by the optical density of the samples measured at 600 nm ($OD_{600}$).

Then, ME and ME/Ag foams (weight of each foam c.a. 10 mg) with a cylindrical shape were employed as filters in a continuous-flow setup. The filtration area and thickness of the foams were 0.5 cm$^2$ and 2 cm, respectively. Filtration tests were carried out at a flow rate of 100 mL/h·cm$^2$, using 50 mL of an *E. coli* suspension (100,000 CFU/mL in milliQ sterile water) as feeding liquid. The filtration process lasted 1 hour, and the output water was collected. Three portions of 15 mL of the filtered water were taken from the collected water for further analysis. Water portions of 1 mL were also collected at random times during the filtration process, in order to perform the antibacterial tests. In all cases, 200 µL aliquots of the filtered liquids were immediately plated as such (undiluted), and after 1:10 dilution with fresh medium (*i.e.*, milliQ sterile water). Plates were stored at 37° C overnight, allowing to the bacteria to grow.

**3. Results and Discussion**

3.1 *In situ* synthesis of Ag NPs on the ME foams

As a consequence of the treatment of the ME foams with the $AgNO_3$ solutions, the samples presented an intense color change, from light gray (pristine ME foams, **Figure 1A**) to brown (treated ME/Ag foams, **Figure 1B**), a clear indication of the presence of Ag NPs.[10,17,26,28,29] The increase of the concentration of the $AgNO_3$ solution, from 0.05 to 1.70 mg/mL, and/or of the reaction time between the ME and the precursor solution, from 1 to 7 days, led to a darker brown color of the foams (see Supporting Information, Figure S2), indicating the increase of the amount of the formed Ag NPs. To confirm that the treatment procedure allows to control the amount of silver transferred to the ME foams, ICP analysis of the ME/Ag foams dipped at precursor solutions of different concentrations and for different time intervals was performed. As shown in **Figure 1C**, the silver content of the treated ME foams can be easily controlled, starting from 0.2 wt.% of silver in the ME foam dipped for 1 day in a 0.05 mg/mL solution to 18.6 wt.% of silver after dipping in a 1.70 mg/mL solution for 7 days. ME/Ag foams treated with solutions of $AgNO_3$ concentrations of 0.05 or 0.10 mg/mL reach a plateau after 5 or 7 days of dipping, respectively. On the contrary, it should be noticed that higher Ag contents on the foams can be obtained by increasing the concentrations of the precursor solutions (0.85 and 1.70 mg/mL) and the reaction times, as no plateau has been reached even after 7 days of treatment. Despite that, the maximum value of silver transferred to the foam achieved in this work (18.6 wt.%) is comparable to the best results reported up to date in terms of silver transfer to a polymeric matrix (25.6 wt.% for cotton fibers, see **Table 1**), and the highest achieved following a simple single-step procedure (previous reported up to 0.3 wt.%, see Table 1).[8,10,11,14,15,18,19,30–32]

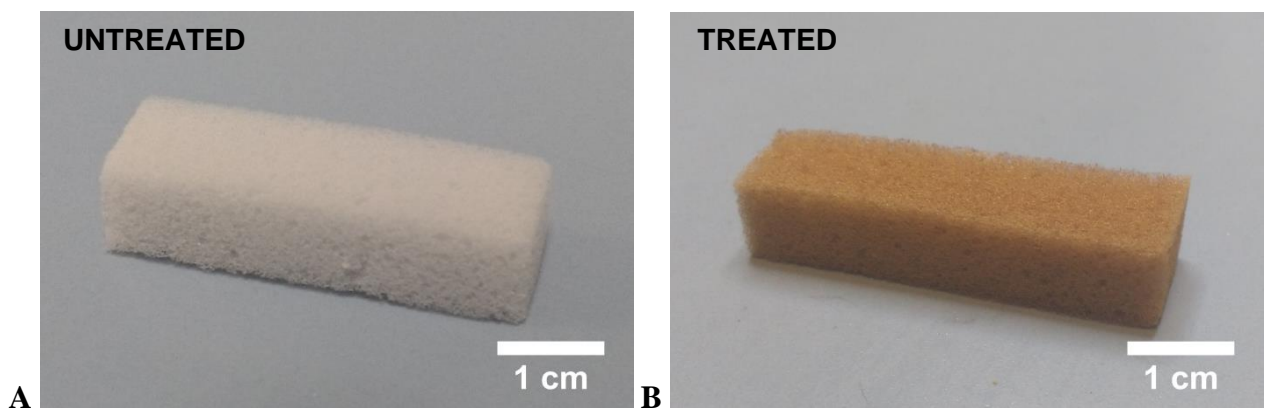

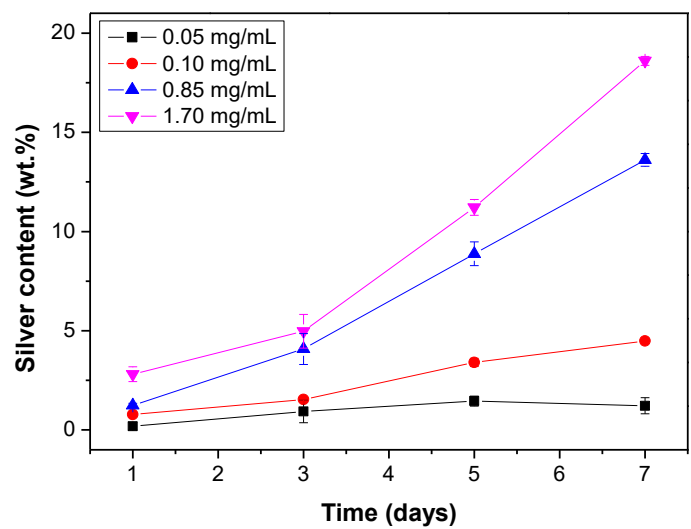

C

**Figure 1.** Photographs of the untreated ME (A) and treated (B) ME/Ag foams. Silver content (wt.%) measured by ICP in the treated ME/Ag foams, obtained from different AgNO$_3$ concentrations, as a function of the reaction time (C).

**Table 1.** Porous nanocomposite materials obtained by Ag NPs incorporation and their maximum silver content achieved with respect to the total weight of the nanocomposite (wt.%).

| *Material* | *Procedure* | *Method employed to transfer the Ag NPs* | *Silver (wt.%)* | *Ref.* |
|---|---|---|---|---|
| Latex foam | Single-step | Dip-coating of the foams in a Ag NPs solution | 0.3 | [14] |
| Polyurethane foam | Single-step | Dip-coating of the foams in a Ag NPs solution | 0.3 | [15] |
| Commercial textiles | N/A | N/A | 2.2 | [32] |
| Paper | Multiple-step | Impregnation of the paper with AgNO$_3$ and subsequent *in situ* synthesis by immersion in a NaBH$_4$ solution | 2.7 | [10] |
| Silk fibers | Multiple-step | Coating of silk fibers with polydopamine and subsequent *in situ* synthesis by immersion in a $AgNO_3$ solution | 4.0 | [18] |
| Polypropylene fibers | Single-step | Mixing of silver nanopowder and polypropylene by extrusion | 5.0 | [30] |
| Cotton fibers | Multiple-step | Immersion of the fabric in a $AgNO_3$ solution of water and ethylene glycol and subsequent *in situ* synthesis by irradiation and addition of ammonia | 6.0 | [19] |
| Polyimide fibers | Multiple-step | Mixing of silver acetate with polyimide, production of the fibers, and thermally-induced *in situ* synthesis | 7.0 | [21] |
| ME foam | Multiple-step | Immersion of the foams in a dopamine hydrochloride solution, immersion of the foams into a graphene oxide (GO)/$Ag^+$ solution, and subsequently in a $NaBH_4$ solution for the *in situ* synthesis of the Ag NPs | 9.8 | [8] |
| **ME foam** | **Single-step** | ***In situ* synthesis on the ME foam by immersion in a $AgNO_3$ solution** | **18.6** | **This work** |
| Cotton fibers | Multiple-step | Treatment of the cotton fibers with a NaOH solution, immersion in an $[Ag(NH_3)_2]^+$ solution and subsequent *in situ* synthesis by immersion in a glucose solution | 25.6 | [11] |

The synthesis of the Ag NPs on the struts surface of ME/Ag foams did not modify their porous structure, as the same morphology was observed by a low magnification SEM analysis (**Figures 2A and 2B**). In particular, smooth strut surfaces similar to the pristine foams were observed. However, higher magnification images show the presence of the Ag NP aggregates on the surface of the struts of the ME/Ag foams, as clearly shown in the inset of Figure 2B and **Figure 2C**. Although the present imaging process did not allow the identification of individual Ag NPs on the ME/Ag foams, a homogeneous distribution of silver on the struts surface of ME/Ag foams was confirmed by the EDS analysis (**Figure 2C**).

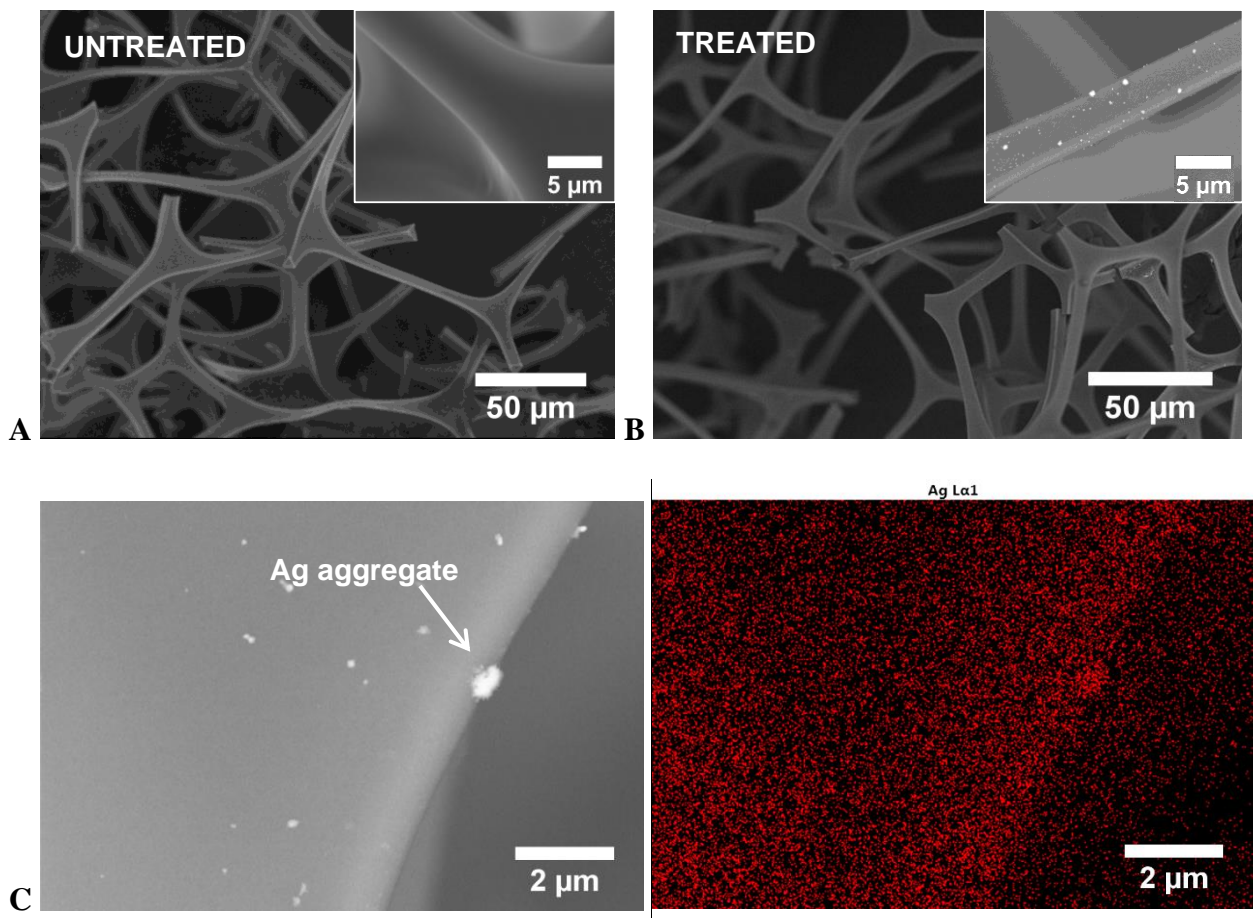

**Figure 2.** SEM micrographs of the untreated ME (A) and treated ME/Ag foam with 1.70 mg/mL AgNO$_3$ solution for 3 days (11.2 wt.% of silver was transferred to the foam) (B). SEM image of the ME/Ag foam and the corresponding EDS map showing a homogeneous distribution of silver (red) (C).

Further analysis of the ME/Ag foam surfaces by TEM was necessary for a direct observation of the individual Ag NPs (**Figure 3**). As shown at the cross section TEM image, small Ag NPs with spherical shape and sizes between 2 and 20 nm (average size ~ 7.2 ± 3.6 nm, inset of Figure 2) were distributed on the ME struts surface even after the shortest reaction times (1 day). It should be noticed that no NPs were detected inside the struts, indicating that the process can precisely and exclusively modify the surface of the pores.

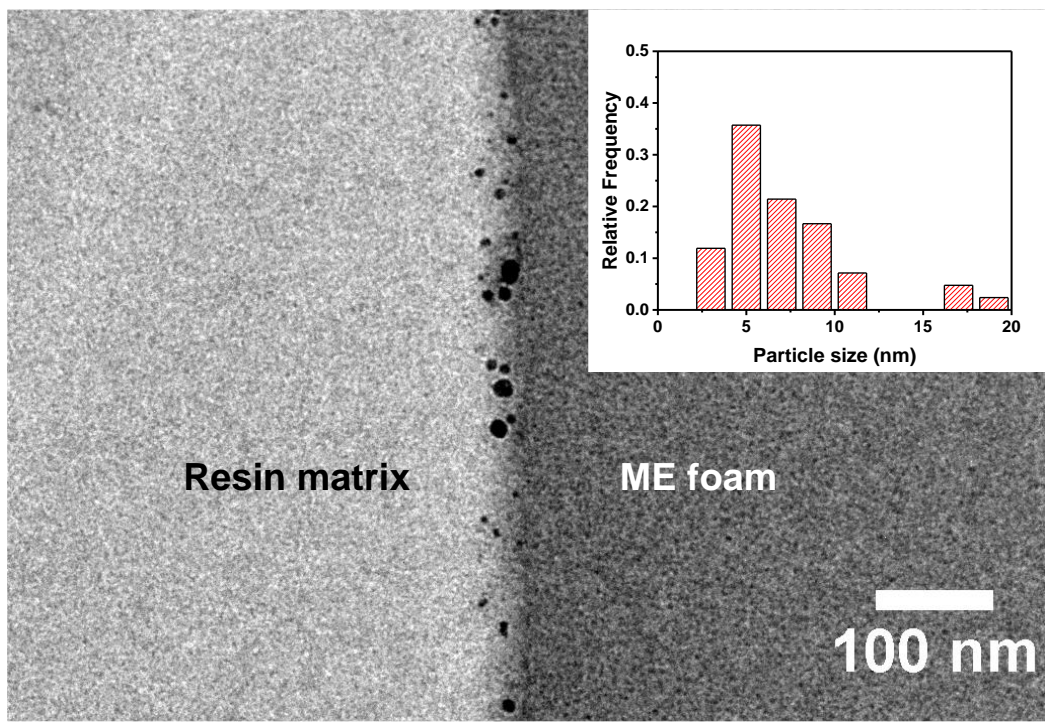

**Figure 3.** TEM micrograph of the surface of a ME/Ag foam treated with 0.85 mg/mL of AgNO$_3$ solution for 1 day. The light gray area (left) corresponds to the resin matrix in which the foam was imbibed, whereas the dark gray area (right) corresponds to a strut of the ME foam. Size distribution of the Ag NPs is shown in the inset.

The observed *in situ* synthesis of Ag NPs on the surfaces of the struts of the ME foams directly from the AgNO$_3$ solution, without any other chemical or radiation process involved, has not been previously described in the literature, to the best of our knowledge. Therefore, to identify the processes involved in the synthesis of the NPs, further chemical characterization was performed. In particular, the formation of Ag NPs on the ME foam from the silver precursor solution was chemically characterized by XPS (**Figure 4**).

**Figure 4A** shows the wide spectra of pristine ME foams and ME/Ag foams after 3 days immersion in an AgNO$_3$ solution at 1.7 mg/mL (XPS results were equivalent at any concentration or reaction time; spectra with best signal to noise ratio are shown here). Nitrogen, carbon, and oxygen from ME polymer chains are clearly identified in the pristine sample by XPS peaks (N 1s, C 1s, and O 1s) as well as by X-ray-induced Auger peaks (O KLL). Additionally, sodium and sulfur are also identified (Na 1s, S 2p, and Na KLL peaks) and can be ascribed to sodium bisulfite, which is a common additive of ME foams.[33] After treatment, the presence of silver in the sample is clearly evidenced by the Ag 3d and Ag 3p peaks. To gain more insight into the chemical reactions happening during the treatment, and especially to determine whether the silver signal comes from metallic NPs or unreacted residual precursors, high-resolution XPS spectra of selected peaks were acquired (**Figure 4B**). The Ag 3d spectrum of the treated foam was fitted with two doublets corresponding to metallic silver (Ag $3d_{5/2}$ at binding energy 367.9 +/- 0.1 eV) thus confirming the presence of $Ag^0$ NPs on the ME foams, and residues of Ag(I) (Ag $3d_{5/2}$ at binding energy 368.9 +/- 0.1 eV) deriving from the $AgNO_3$ precursor.[34] It was found that the relative amount of the Ag(I) compared to $Ag^0$ (Ag(I) / (Ag(I) + $Ag^0$)) was only about 13%. Other ME/Ag foams obtained using different $AgNO_3$ concentrations or reaction times showed similar or slightly lower relative amounts of Ag(I) compared to the $Ag^0$. These results are indicative of a reduction of silver from $Ag^+$ to $Ag^0$ on the surface of the foams. The evolution of N 1s spectrum suggests that such reduction is allowed by an oxidation of chemical groups on the melamine chains, as explained hereafter. The N 1s signal of the pristine ME foam was fitted to two components at BE = 399.2 +/- 0.1 eV and BE = 398.0 +/- 0.1 eV representative for the C=*N*-C (hereafter simply *N*=C) and C-*N*H-C or C-$NH_2$ (hereafter simply *N*-C) groups of the ME, showing areas with the expected 1:1 ratio for pure ME.[35] Furthermore, a weak and broad feature around 404 eV was observed amounting for *ca.* 2.5 % of the total N 1s signal (area ratio), ascribed to π-π* interactions of the N atoms in the aromatic ring.[35] On the treated sample, significant differences appear on the N 1s spectrum. First, the *N*-C / *N*=C ratio is lowered to about 0.8. Second, the peak at higher BE is broader and has a slightly larger area (*ca.* 5% of the total N 1s signal), which suggests the formation of one or more new peaks around 406 eV (signal is too low for a precise fit). This new feature can be ascribed to the formation of $NO_x$ chemical groups.[36] Even though these may partly be due to $NO_3^-$ anions coming from the $AgNO_3$ precursor, it is likely that they are also the result of a partial oxidation of melamine chemical groups, which is consistent with the above mentioned reduction in the area of the C-$NH_x$ component. Hence, we conclude that the almost full reduction of Ag(I) to metallic $Ag^0$ NPs at the surface of melamine foams is accompanied by a partial oxidation of C-$NH_x$ to C-$NO_y$ groups of melamine to complete the redox cycle.

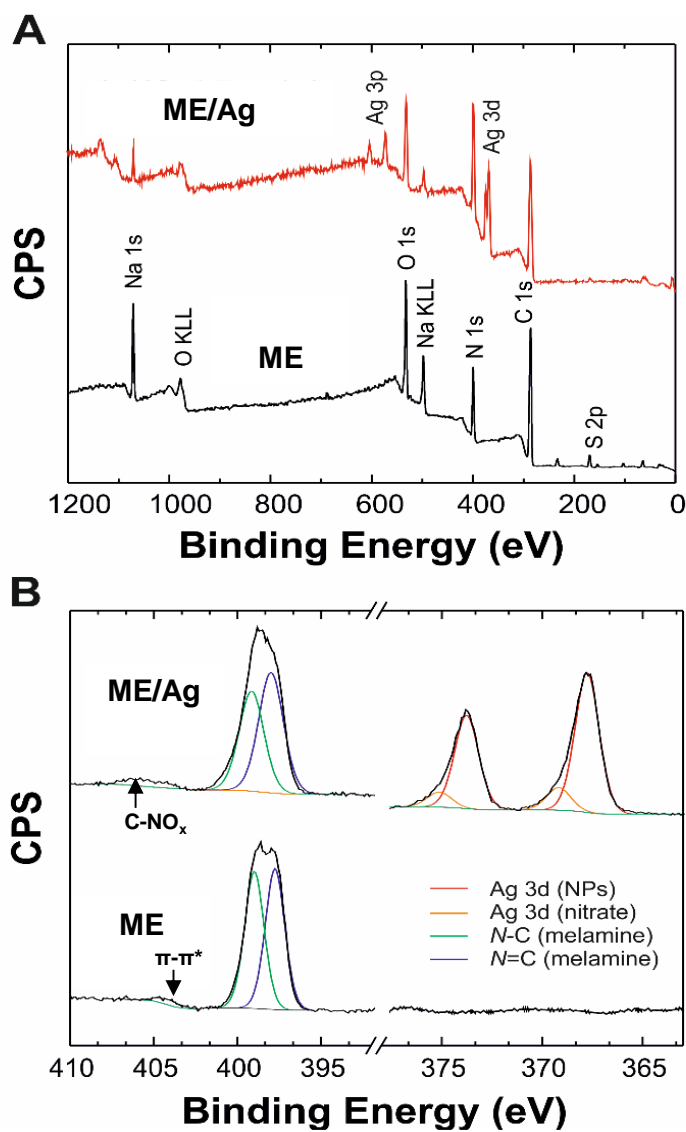

**Figure 4.** XPS wide spectra of pristine ME and ME/Ag treated with a 1.7 mg/mL $AgNO_3$ solution for 3 days (A). High-resolution N 1s and Ag 3d peaks and their deconvolution (B).

To further analyze the chemical modifications induced by the formation of the Ag NPs on the ME foam, ATR-FTIR spectroscopy was carried out (**Figure 5**), demonstrating significant spectroscopic differences between the treated and non-treated foams. In particular, after the dipping of the foams in the $AgNO_3$ solutions, a reduction of the intensity of the shoulder at 3395 cm$^{-1}$, associated with the stretching mode of free N-H, was observed. This indicates that more H-bonds are formed on the ME foams.[37] On the contrary, after the treatment, the increase of the intensity of the vibration band at 1289 cm$^{-1}$, ascribed to the O-H in-plane bending in oxime (=N-OH) functional groups, is observed.[38] Such oximes can be formed by oxidation of primary amines (-NH$_2$) present in the pristine ME foam. Their formation justifies the lower ratio C-N/C=N and the extension of the H-bond network found by XPS and FTIR, respectively, after the treatment of the ME samples. Finally, the band at 810 cm$^{-1}$ observed at the pristine ME was shifted to 804 cm$^{-1}$ for the ME/Ag foams. This peak is attributed to a bending mode of the triazine ring and the shift can be related to changes in the aromaticity of the ring induced by the formation of the oximes.[39] It should be noticed that all the observed spectral changes are more intense at the foams treated with precursor solutions of higher concentration. This indicates a more efficient chemical interaction between the AgNO$_3$ solution and the ME foams, in good agreement with the results obtained so far, which prove the higher amount of silver transferred to the foams after their treatment with solutions of higher AgNO$_3$ concentrations (Figure 1C).

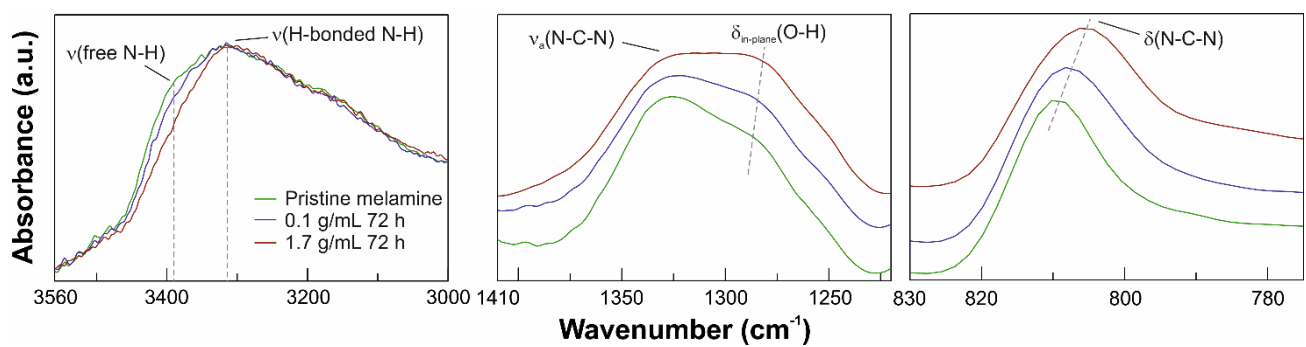

**Figure 5.** ATR-FTIR spectra of pristine ME and ME/Ag foams treated with AgNO$_3$ solution at 0.1 and 1.7 mg/mL for 3 days. The spectral ranges investigated are between 3560-3000 cm$^{-1}$, 1410-1190 cm$^{-1}$, and 830-760 cm$^{-1}$. The assignment of the bands is included in each case.

Finally, the employed AgNO$_3$ solutions were also studied before and after the reaction with the ME foams. **Figure 6** compares the UV-vis spectra of a AgNO$_3$ solution at 0.1 mg/mL before and after 1 day of immersion of a ME foam (results were similar at any concentration or reaction time, but the selected conditions provided a higher signal to noise ratio). Before the reaction with the ME foam, the NO$_3^-$ ions in the solution showed two peaks at ~277 and ~285 nm. These positions are typical of n → π* transits for NO$_3^-$ ion complexes in solution.[40,41] On the other hand, after the reaction, a new contribution in the nitrate region was observed as a shoulder near 300 nm, which is related to NO$_3^-$ anions deriving from nitric acid (*i.e.*, NO$_3^-$ ions not forming complexes with Ag$^+$ ions).[40,41]

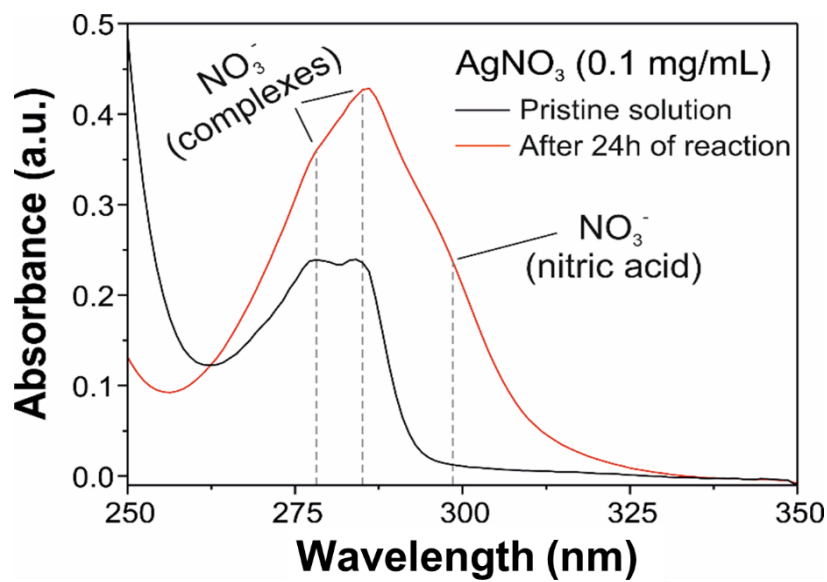

**Figure 6.** UV-vis spectra of a AgNO₃ solution at 0.1 mg/mL before and after 1 day in contact with a ME foam. The assignment of the peaks is included.

Considering the main findings of the above described chemical characterization, which are the presence of metallic silver and oxime groups on the foams, and the formation of nitric acid in the precursor solution, the following redox reaction between the AgNO₃ and the ME polymer is proposed in a H₂O:THF (1:19 v:v) mixture:

Reduction reaction: $(Ag^+ + 1\ e^- \rightarrow Ag^0)$ x3

Oxidation reaction: Melamine-NH$_2$ + H$_2$O → Melamine=N-OH + 3 H$^+$ + 3 e$^-$

Redox reaction (considering the diluted state of the precursor):

3 Ag$^+$NO$_3^-$ + Melamine-NH$_2$ + H$_2$O → 3 Ag$^0$ + Melamine=N-OH + 3 H$^+$NO$_3^-$

The formation of three moles of metallic silver for each mole of amine groups in the ME foam can explain the high values of Ag content (see Figure 1C) obtained with this system.

3.2 Antibacterial performance of the ME/Ag foams

Several works established that the antibacterial performance of a material loaded with Ag NPs or ions is mainly related to its capability to release Ag$^+$ ions to the medium,[4,6–8] and concentrations of about 0.1-1.0 ppm should be enough to provide the antibacterial effect.[2,4,6,10] For this reason, the Ag$^+$ ions release in distilled water from the ME/Ag foams was studied. As shown in **Figure 7A,** the amount of the released Ag$^+$ ions reached a plateau after 10 h of dipping in distilled water, independently of the initial Ag content in the foam. Moreover, the Ag$^+$ ions release was found to be directly proportional to the amount of the silver transferred to the foams. In this sense, ME/Ag foams with 1.8 wt.% silver content were able to release up to 6.6 ppm of Ag$^+$, while ME/Ag foams with 5.5 and 9.6 wt.% silver content released up to 47.9 and 79.0 ppm, respectively (**Figure 7B**). The concentration of the Ag$^+$ released by the ME/Ag foams with Ag contents below 10 wt.% was much higher than the safety threshold for drinking water (0.1 ppm)[4] and also far above the Ag$^+$ concentrations needed for an effective antibacterial performance. Since it is expected that ME/Ag foams with initial Ag contents higher than 10 wt.% with respect to the foam would release even higher quantities of Ag$^+$, such samples were omitted from the antibacterial study that is presented below.

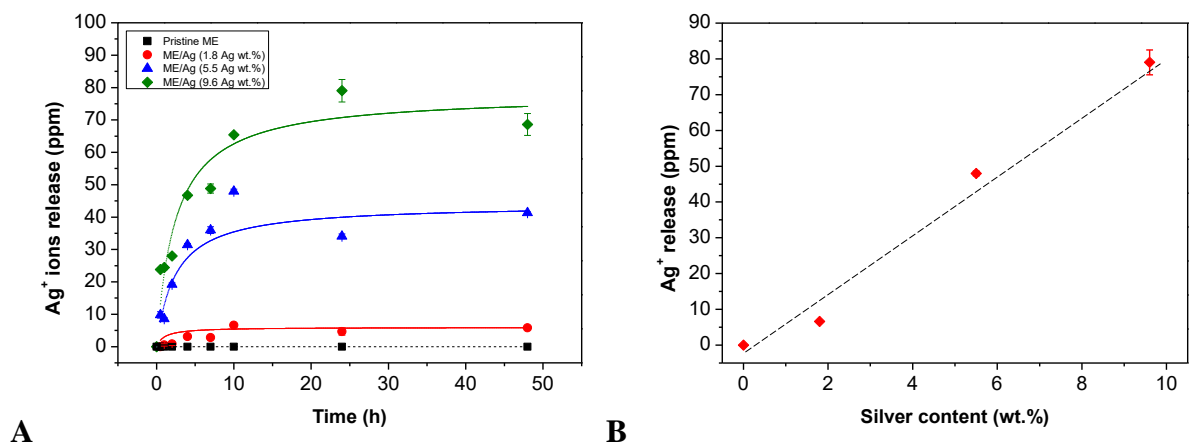

**Figure 7.** Dependence of the Ag$^+$ ions release to the immersion time of the ME/Ag foams in distilled water, for foams with different silver contents (A). Dependence of the released Ag$^+$ ions at plateau to the Ag content of the ME/Ag foams (B). Lines serve as a guide for the eyes.

To confirm that the Ag release is induced exclusively by the ions directly released by the ME/Ag foams, and not by any Ag NPs dispersed in the liquid upon the foams dipping, before the ICP analysis and after the dipping process in the distilled water, the medium was analyzed by UV-vis spectroscopy. As shown in Figure S3, independently of the type of ME/Ag foam employed, no characteristic absorbance peak of the Ag NPs was observed, indicating that the Ag NPs are stably adhered on the ME foams and that the signal of the ICP analysis is mainly attributed to the released ions.

To explore the utilization of the developed foams for water sanitation applications, tube tests were performed by placing the foam samples in a bacterial suspension. As shown in **Figure 8**, regardless of their Ag content, the ME/Ag foams completely hindered the bacterial growth. Hence, even ME/Ag foams with low Ag content (*e.g.*, 1.8 wt.%, which after 1 hour release just 0.5 ppm of Ag$^+$, reaching 3.1 ppm after 4 hours, see Figure 7.B) were able to completely inhibit the bacterial growth, showing constant and low OD$_{600}$ values for the time period of 4 hours. On the contrary, the pristine ME foams did not inhibit the bacterial growth, as shown by the increasing OD$_{600}$ value (Figure 8).

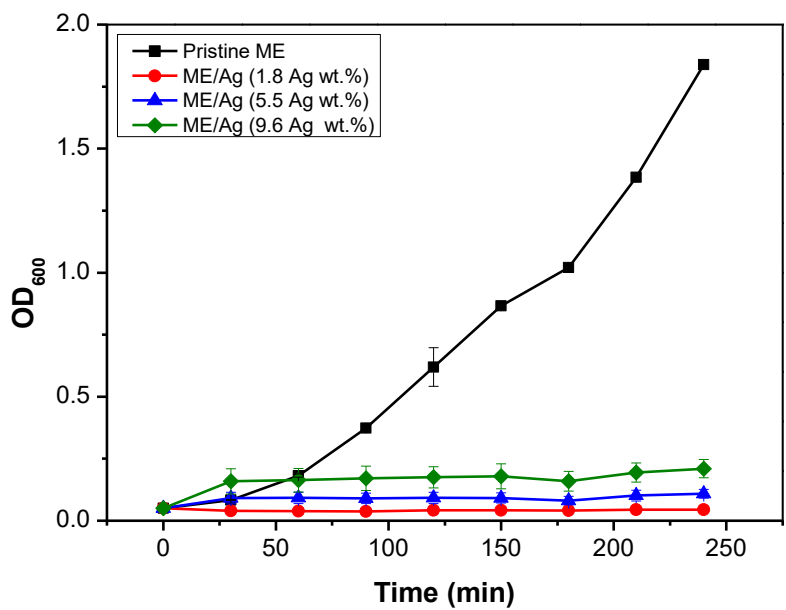

**Figure 8.** Bacterial growth up to 4 hours of the tube tests carried out with ME foams without (Pristine ME) and with Ag NPs (ME/Ag with Ag contents of 1.8, 5.5, and 9.6 wt.%). The slight displacement of the baseline found with the increase of the Ag content is not related to the accidental release of Ag NPs (Figure S3).

Next, the ME/Ag foams were employed as filters in a continuous-flow setup (see Experimental section), where bacteria contaminated water was the feeding medium with a flow rate of 100 mL/h·cm$^2$. To do so, as a first step, the Ag$^+$ release from the ME/Ag foams in the collected water after 1 hour of filtration at such conditions was studied. It was found that ME/Ag foams with low Ag content (1.8 wt.%) presented a Ag$^+$ release below the ICP detection limit of the performed measurements (0.01 ppm), while ME/Ag foams with medium Ag content (9.6 wt.%) showed a suitable Ag$^+$ release, with an Ag$^+$ concentration in the collected water slightly below 1 ppm. Taking into account that Ag$^+$ concentrations of about 0.1-1.0 ppm are enough to provide an antibacterial effect, ME/Ag foams with Ag content of 9.6 wt.%. were utilized in the specific test.

As expected, the pristine ME foams did not show any antibacterial effect, finding significant bacterial growth in the collected liquid after filtration (**Figure 9A**). On the contrary, the ME/Ag foams showed a clear antibacterial effect, with no bacterial growth in the collected water after 1 hour of filtration (**Figure 9B**). Same results were found when the fractions of the collected water after 1 hour of filtration were diluted with fresh media (1:10), to ensure that the $Ag^+$ concentration is below the safety threshold (< 0.1 ppm) for drinkable water (**Figure 9C**).

To define the kind of antibacterial behavior achieved, by identifying the influence of the contact time between the bacteria and the $Ag^+$ ions on the antibacterial effect, these results were compared with the ones obtained when filtered water was collected from the filtration setup output at random times during the filtration process. In order to ensure that the concentration of the $Ag^+$ ions released by the foams during the dynamic process of filtration is above the limit for antibacterial activity, a preliminary ICP study at the collected liquids was performed. It was found that the liquid collected directly from the filtration setup output after 15 minutes from the beginning of the filtration process, contained $Ag^+$ ions with concentration of 2 ppm, while after 30 minutes from the beginning of the filtration process this concentration was below 0.5 ppm, lower than the concentration found in the collected liquid after 1 hour filtration. Thus, in order to have a similar or higher $Ag^+$ ions concentration than those of the collected liquid after 1 hour, the liquid collected from the filtration output after 15 minutes from the beginning of filtration was further analyzed. As shown in **Figure 9D,** no bacteria grew in the undiluted fractions. However, after dilution (1:10) with fresh media (**Figure 9E**), these fractions provided a comparable bacterial growth with the control samples (**Figure 9F**). Same results would be also expected for the liquids collected from the filtration output after various times from the beginning of the filtration process, with only limitation the $Ag^+$ ions concentration to be higher than the defined limit for antibacterial activity.

Therefore, water samples immediately collected after passing through the foams showed a bacteriostatic behavior, since as proved, the bacterial growth is inhibited only when the $Ag^+$ concentration presents certain values. This behavior is lost once the $Ag^+$ concentration decreases, probably due to the short contact time between $Ag^+$ ions and the bacteria. Indeed, in such process the bacteria are in contact with the $Ag^+$ ions of concentration over the safety threshold for a few minutes before the dilution with the fresh medium. On the contrary, the water samples taken from the collected water after 1 h presented a bactericide behavior, since the bacteria were killed and cannot growth again even if the $Ag^+$ concentration decreases. This can be attributed to the longer contact time between the $Ag^+$ ions at a concentration over the safety threshold and the bacteria, which allows not only to hinder temporarily the bacteria proliferation, but also to kill them before dilution with the fresh liquid medium. Accordingly, these ME/Ag foams could be actually employed for the treatment of water polluted with bacteria, as a complete removal of the bacteria can be achieved just by collecting the filtered water before the further release of the treated water and mixing with fresh water.

The flow rate successfully tested, 100 mL/h·cm$^2$, is more than three times higher than the flow rates previously employed in the literature (~30 mL/h·cm$^2$) following a similar approach (*i.e.*, using a polymer foams coated with Ag NPs as water filters). Specifically, as suggested by Jain and Pradeep,[16] the use of this approach in combination with conventional ceramic water filter candles allows to reach typical flow rates as the ones used in domestic water purifiers (0.5 L/min). In particular, with the herein proposed ME/Ag foams, a flow rate up to 1.6 L/min can be reached, considering a typical surface area of one of these filters (~1000 cm$^2$), indicating that they could work properly in combination with such ceramic water filters.

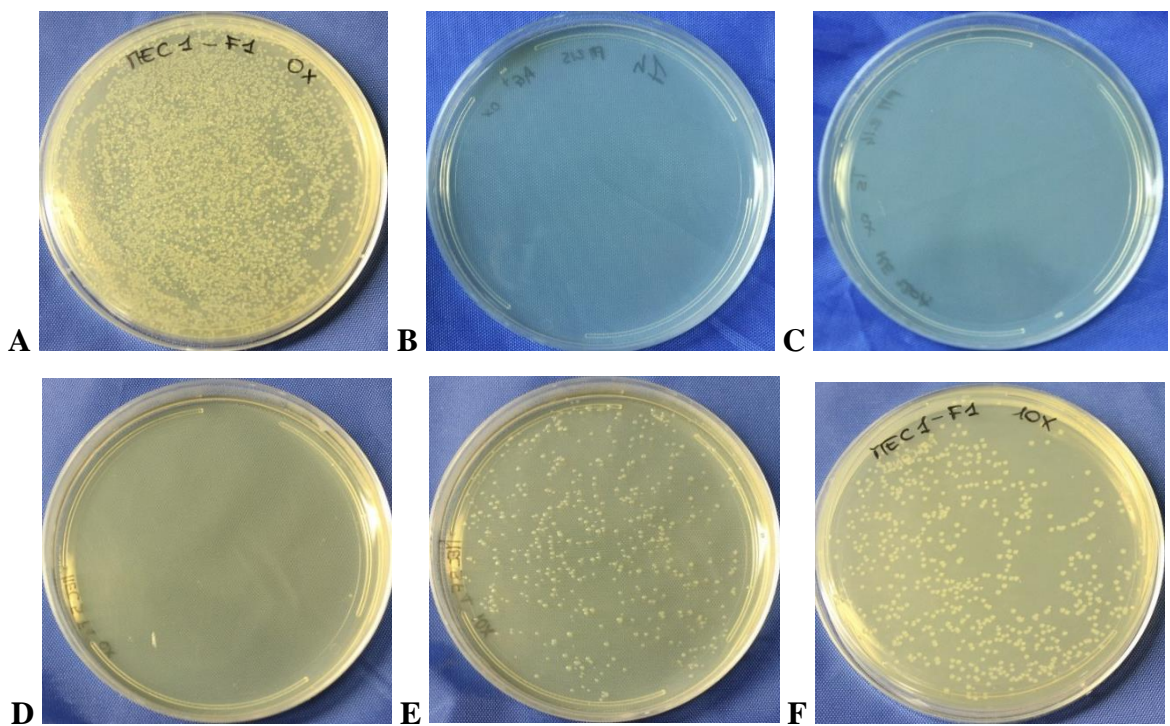

**Figure 9.** Continuous flow tests performed at 100 mL/h·cm$^2$ using two types of foams ME and ME/Ag with 9.6 wt.% of silver content. Bacterial growth obtained from the collected water after 1 h of filtration using the ME (A), and the ME/Ag foam (B). Bacterial growth of the collected water after 1 h filtration and 1:10 dilution with a fresh medium using the ME/Ag foam (C). Bacterial growth of the water taken directly from the filtration output after 15 minutes filtration (D), and after subsequent 1:10 dilution with a fresh medium (E) using the ME/Ag foam, and a control sample (ME foam) (F).

In view of the above results, the developed ME/Ag foams are suitable candidates for the treatment of bacteria-polluted water. Moreover, their notable bactericide performance in combination with the excellent mechanical properties, which are not modified after the treatment, allow also their use in other applications, such as the production of antibacterial cushions and mattresses (see Supporting Information). It should be noticed that the simple single-step fabrication process of the ME/Ag foams can be easily scaled up, and does no present any limitations regarding the geometry or dimensions of the treated foams (see Supporting Information, Figure S6). In addition, the proposed procedure allows fine control of the silver transferred to the ME foams (see Figure 1C), as well as the achievement of the highest amount in the literature by a single-step method (see Table 1). Therefore, the proposed ME/Ag foams are promising candidates for the facile production of porous materials with remarkable antibacterial performance.

**Conclusions**

This work proposes a single-step route to produce melamine foams decorated with Ag NPs (ME/Ag foams) formed *in situ* directly onto the surface of struts of the foams. The presence of Ag NPs with sizes below 10 nm uniformly distributed on the struts has been demonstrated by electron microscopy studies, whereas the synthesis reaction mechanism of the NPs has been proposed. In particular, it has been proved that the R-NH$_2$ groups of the ME foams are able to react with the AgNO$_3$ precursor solution, inducing the reduction of the Ag$^+$ ions into Ag NPs. The amount of Ag NPs transferred to the ME foams is finely controlled by the concentration of the silver precursor solution and the dipping time, allowing to reach silver contents ranging from 0.2 wt.% up to, at least, 18.6 wt.%. It should be highlighted that the maximum silver content achieved in this work (*i.e.,* 18.6 wt.%) is, to the best of our knowledge, the highest silver content transferred to polymer foam by a single-step procedure.

The proved stable adhesion of the Ag NPs to the ME foams allowed their use in water treatment, as only Ag$^+$ ions, and not Ag NPs, are released in contact with the water. The obtained ME/Ag foams have shown excellent antibacterial performance against *E. Coli* in static (tube test) and continuous-flow (filtration) tests. Moreover, it has been proved that these ME/Ag foams could provide a bactericide behavior in filtration applications, even when the treated water is further diluted with fresh water, to ensure that the Ag$^+$ ions concentration is below the safety threshold for drinking water (< 0.1 ppm). These unique features make these flexible ME/Ag foams suitable candidates for their use in water sanitation by filtration of bacteria-polluted water.

**Acknowledgments**

The authors acknowledge Tiziano Catelani (Electron Microscopy Facility, Istituto Italiano di Tecnologia) for performing the TEM analysis, and Francesca Mussino (Nanobiointeractions & Nanodiagnostics, Istituto Italiano di Tecnologia) for her assistance with the antibacterial tests. The Materials Characterization Facility at the Istituto Italiano di Tecnologia is also acknowledged for help with XPS characterization.

**Supporting Information.** UV-Vis spectra of the treated water after being in contact with the ME/Ag foams. Percentage atomic concentrations calculated by the survey XPS spectra for all species present on the ME samples before and after the treatment with AgNO$_3$. Photographs of ME/Ag foams showing the color change due to the presence of the formed Ag NPs. Additional considerations and results about the potential use of the ME/Ag foams as antibacterial cushions and mattresses. Photographic evidences of the production of ME/Ag foams with different geometry and large dimensions.

**References**

(1) (WHO), W. H. O. Drinking-Water http://www.who.int/mediacentre/factsheets/fs391/en/.

(2) Rai, M.; Yadav, A.; Gade, A. Silver Nanoparticles as a New Generation of Antimicrobials. *Biotechnol. Adv.* **2009**, *27* (1), 76–83.

(3) Prabhu, S.; Poulose, E. K. Silver Nanoparticles: Mechanism of Antimicrobial Action, Synthesis, Medical Applications, and Toxicity Effects. *Int. Nano Lett.* **2012**, *2*, 32.

(4) Health, C. for R. into E. and. *Silver: Water Disinfection and Toxicity*; World Health Organization, 2014.

(5) Silvestry-Rodriguez, N.; Sicairos-Ruelas, E. E.; Gerba, C. P.; Bright, K. R. Silver as a Disinfectant. In *Rev Environ Contam Toxicol.*; 2007; Vol. 191, pp 23–45.

(6) Reidy, B.; Haase, A.; Luch, A.; Dawson, K.; Lynch, I. Mechanisms of Silver Nanoparticle Release, Transformation and Toxicity: A Critical Review of Current Knowledge and Recommendations for Future Studies and Applications. *Materials (Basel).* **2013**, *6* (6), 2295–2350.

(7) Rizzello, L.; Pompa, P. P. Nanosilver-Based Antibacterial Drugs and Devices: Mechanisms, Methodological Drawbacks, and Guidelines. *Chem. Soc. Rev.* **2014**, *43* (5), 1501–1518.

(8) Deng, C. H.; Gong, J. L.; Zhang, P.; Zeng, G. M.; Song, B.; Liu, H. Y. Preparation of Melamine Sponge Decorated with Silver Nanoparticles-Modified Graphene for Water Disinfection. *J. Colloid Interface Sci.* **2017**, *488*, 26–38.

(9) De Matteis, V.; Malvindi, M. A.; Galeone, A.; Brunetti, V.; De Luca, E.; Kote, S.; Kshirsagar, P.; Sabella, S.; Bardi, G.; Pompa, P. P. Negligible Particle-Specific Toxicity Mechanism of Silver Nanoparticles: The Role of Ag+ Ion Release in the Cytosol. *Nanomedicine Nanotechnology, Biol. Med.* **2015**, *11* (3), 731–739.

(10) Dankovich, T. A.; Gray, D. G. Bactericidal Paper Impregnated with Silver Nanoparticles for Point-of-Use Water Treatment. *Environ. Sci. Technol.* **2011**, *45* (5), 1992–1998.

(11) Xue, C.-H.; Chen, J.; Yin, W.; Jia, S.-T.; Ma, J.-Z. Superhydrophobic Conductive Textiles with Antibacterial Property by Coating Fibers with Silver Nanoparticles. *Appl. Surf. Sci.* **2012**, *258* (7), 2468–2472.

(12) Mikelonis, A. M.; Lawler, D. F.; Passalacqua, P. Multilevel Modeling of Retention and Disinfection Efficacy of Silver Nanoparticles on Ceramic Water Filters. *Sci. Total Environ.* **2016**, *566–567*, 368–377.

(13) Rayner, J.; Luo, X.; Schubert, J.; Lennon, P.; Jellison, K.; Lantagne, D. The Effects of Input Materials on Ceramic Water Filter Efficacy for Household Drinking Water Treatment. *Water Sci. Technol. Water Supply* **2017**, *17* (3), 859–869.

(14) Song, J. Y.; Kwon, E.-Y.; Kim, B. S. Antibacterial Latex Foams Coated with Biologically Synthesized Silver Nanoparticles Using Magnolia Kobus Leaf Extract. *Korean J. Chem. Eng.* **2012**, *29* (12), 1771–1775.

(15) Phong, N. T. P.; Thanh, N. V. K.; Phuong, P. H. Fabrication of Antibacterial Water Filter by Coating Silver Nanoparticles on Flexible Polyurethane Foams. *J. Phys. Conf. Ser.* **2009**, *187*, 12079.

(16) Jain, P.; Pradeep, T. Potential of Silver Nanoparticle-Coated Polyurethane Foam as an Antibacterial Water Filter. *Biotechnol. Bioeng.* **2005**, *90* (1), 59–63.

(17) Rathnayake, W. G. I. U.; Ismail, H.; Baharin, A.; Darsanasiri, A. G. N. D.; Rajapakse, S. Synthesis and Characterization of Nano Silver Based Natural Rubber Latex Foam for Imparting Antibacterial and Anti-Fungal Properties. *Polym. Test.* **2012**, *31* (5), 586–592.

(18) Lu, Z.; Xiao, J.; Wang, Y.; Meng, M. In Situ Synthesis of Silver Nanoparticles Uniformly Distributed on Polydopamine-Coated Silk Fibers for Antibacterial Application. *J. Colloid Interface Sci.* **2015**, *452*, 8–14.

(19) Perelshtein, I.; Applerot, G.; Perkas, N.; Guibert, G.; Mikhailov, S.; Gedanken, A. Sonochemical Coating of Silver Nanoparticles on Textile Fabrics (Nylon, Polyester and Cotton) and Their Antibacterial Activity. *Nanotechnology* **2008**, *19* (24), 245705.

(20) Picca, R. A.; Paladini, F.; Sportelli, M. C.; Pollini, M.; Giannossa, L. C.; Franco, C. Di; Panico, A.; Mangone, A.; Valentini, A.; Cioffi, N. Combined Approach for the Development of Efficient and Safe Nanoantimicrobials: The Case of Nanosilver-Modified Polyurethane Foams. *ACS Biomater. Sci. Eng.* **2017**, *3*, 1417−1425.

(21) Zhang, Q.; Wu, D.; Qi, S.; Wu, Z.; Yang, X.; Jin, R. Preparation of Ultra-Fine Polyimide Fibers Containing Silver Nanoparticles via in Situ Technique. *Mater. Lett.* **2007**, *61* (19–20), 4027–4030.

(22) Perera, S.; Bhushan, B.; Bandara, R.; Rajapakse, G.; Rajapakse, S.; Bandara, C. Morphological, Antimicrobial, Durability, and Physical Properties of Untreated and Treated Textiles Using Silver-Nanoparticles. *Colloids Surfaces A Physicochem. Eng. Asp.* **2013**, *436*, 975–989.

(23) Lu, H.; Yu, L.; Liu, Q.; Du, J. Ultrafine Silver Nanoparticles with Excellent Antibacterial Efficacy Prepared by a Handover of Vesicle Templating to Micelle Stabilization. *Polym. Chem.* **2013**, *4* (12), 3448.

(24) Zou, K.; Liu, Q.; Chen, J.; Du, J. Silver-Decorated Biodegradable Polymer Vesicles with Excellent Antibacterial Efficacy. *Polym. Chem.* **2014**, *5* (2), 405–411.

(25) Zhang, F.; Wu, X.; Chen, Y.; Lin, H. Application of Silver Nanoparticles to Cotton Fabric as an Antibacterial Textile Finish. *Fibers Polym.* **2009**, *10* (4), 496–501.

(26) Zhang, G.; Liu, Y.; Gao, X.; Chen, Y. Synthesis of Silver Nanoparticles and Antibacterial Property of Silk Fabrics Treated by Silver Nanoparticles. *Nanoscale Res Lett* **2014**, *9* (1), 216.

(27) Abràmoff, M. D.; Magalhães, P. J.; Ram, S. J. Image Processing with ImageJ Second Edition. *Biophotonics Int.* **2004**, *11* (7), 36–42.

(28) Kong, H.; Jang, J. Antibacterial Properties of Novel Poly(methyl Methacrylate) Nanofiber Containing Silver Nanoparticles. *Langmuir* **2008**, *24*, 2051–2056.

(29) Budama, L.; Çakır, B. A.; Topel, Ö.; Hoda, N. A New Strategy for Producing Antibacterial Textile Surfaces Using Silver Nanoparticles. *Chem. Eng. J.* **2013**, *228*, 489–495.

(30) Yeo, S. Y.; Jeong, S. H. Preparation and Characterization of Polypropylene/silver Nanocomposite Fibers. *Polym. Int.* **2003**, *52* (7), 1053–1057.

(31) Dagostin, V. S.; Golçalves, D. L.; Pacheco, C. B.; Almeida, W. B.; Thomé, I. P.; Pich, C. T.; Paula, M. M. S.; Silva, L.; Angioletto, E.; Fiori, M. A. Bactericidal Polyurethane Foam Mattresses: Microbiological Characterization and Effectiveness. *Mater. Sci. Eng. C* **2010**, *30* (5), 705–708.

(32) Geranio, L.; Heuberger, M.; Nowack, B. The Behavior of Silver Nanotextiles during Washing. *Environ. Sci. Technol.* **2009**, *43*, 8113–8118.

(33) Stolz, A.; Le Floch, S.; Reinert, L.; Ramos, S. M. M.; Tuaillon-Combes, J.; Soneda, Y.;

Chaudet, P.; Baillis, D.; Blanchard, N.; Duclaux, L.; San-Miguel, A. Melamine-Derived Carbon Sponges for Oil-Water Separation. *Carbon N. Y.* **2016**, *107*, 198–208.

(34) Ferraria, A. M.; Carapeto, A. P.; Botelho do Rego, A. M. X-Ray Photoelectron Spectroscopy: Silver Salts Revisited. *Vacuum* **2012**, *86* (12), 1988–1991.

(35) Dementjev, A. P.; U, A. de G.; Sanden, M. C. M. van de; Maslakov, K. I.; Naumkin, A. V; Serov, A. A. X-Ray Photoelectron Spectroscopy Reference Data for Identification of the C N Phase in Carbon]nitrogen Films. *Diam. Relat. Mater.* **2000**, *9*, 1904–1907.

(36) Wagner, C. D. *Handbook of X-Ray Photoelectron Spectroscopy: A Reference Book of Standard Data for Use in X-Ray Photoelectron Spectroscopy*; Physical Electronics Division, Perkin-Elmer Corp., 1979.

(37) Bellamy, L. J. The Infra-Red Spectra of Complex Molecules, 3rd Chapman and Hall. London 1975.

(38) Palm, A.; Werbin, H. Infrared Study of the N-OH Group on Alpha and Beta Oximes. *Can. J. Chem.* **1954**, *32* (9), 858–863.

(39) Merline, D. J.; Vukusic, S.; Abdala, A. A. Melamine Formaldehyde: Curing Studies and Reaction Mechanism. *Polym. J.* **2012**, *45* (4), 413–419.

(40) Gvozdić, V.; Tomišić, V.; Butorac, V.; Simeon, V. Association of Nitrate Ion with Metal Cations in Aqueous Solution: A UV-Vis Spectrometric and Factor-Analytical Study. *Croat. Chem. Acta* **2009**, *82* (2), 553–559.

(41) Woodhead, J. L. The Ultra-Violet Absorption Spectra of Nitric Acid and Metal Nitrates in Tri-N-Butyl Phosphate Solutions. *J. Inorg. Nucl. Chem.* **1965**, *27*, 1111–1116.

# TOC graphic

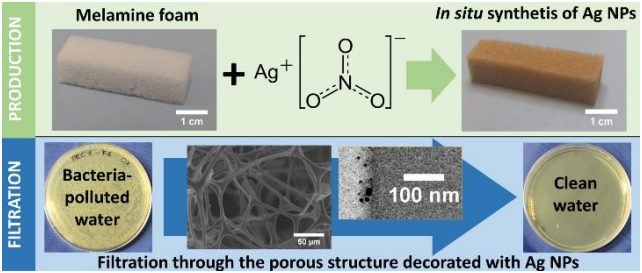